# A Comparison of Water Vapor Line Parameters for Modeling the Venus Deep Atmosphere

**Jeremy Bailey**[*]

School of Physics, University of New South Wales, NSW 2052, Australia

*Corresponding Author E-mail address: j.bailey@unsw.edu.au

# ABSTRACT

The discovery of the near infrared windows into the Venus deep atmosphere has enabled the use of remote sensing techniques to study the composition of the Venus atmosphere below the clouds. In particular, water vapor absorption lines can be observed in a number of the near-infrared windows allowing measurement of the $H_2O$ abundance at several different levels in the lower atmosphere. Accurate determination of the abundance requires a good database of spectral line parameters for the $H_2O$ absorption lines at the high temperatures (up to ~700 K) encountered in the Venus deep atmosphere. This paper presents a comparison of a number of $H_2O$ line lists that have been, or that could potentially be used, to analyze Venus deep atmosphere water abundances and shows that there are substantial discrepancies between them. For example, the early high-temperature list used by Meadows and Crisp (1996) had large systematic errors in line intensities. When these are corrected for using the more recent high-temperature BT2 list of Barber et al. (2006) their value of 45±10 ppm for the water vapor mixing ratio reduces to 27±6 ppm. The HITRAN and GEISA lists used for most other studies of Venus are deficient in "hot" lines that become important in the Venus deep atmosphere and also show evidence of systematic errors in line intensities, particularly for the 8000 to 9500 $cm^{-1}$ region that includes the 1.18 μm window. Water vapor mixing ratios derived from these lists may also be somewhat overestimated. The BT2 line list is recommended as being the most complete and accurate current representation of the $H_2O$ spectrum at Venus temperatures.

# 1   Introduction

Water vapor ($H_2O$) is an important trace gas in the Venus lower atmosphere. It is a contributor to the global greenhouse effect that maintains the very high surface temperature. It also plays a significant part in the chemistry of the lower and middle atmosphere through its involvement in the sulfur oxidation cycle that produces $H_2SO_4$ (Krasnopolsky and Pollack, 1994), and in the active photochemistry above the clouds (Yung and DeMore, 1982). Most of the $H_2O$ is found below the cloud base at ~47 km. In the upper atmosphere the $H_2O$ mixing ratio falls to values ranging from <1 to ~4 ppm (parts per million by volume) with indications of substantial variability (Sandor and Clancy, 2005; Gurwell et al., 2006; Bertaux et al., 2006).

In-situ measurements from the Venera and Pioneer Venus entry probes produced a wide range of sometimes conflicting values for the $H_2O$ mixing ratio below the clouds (von Zahn et al., 1983; Taylor et al., 1997). Some of the results indicated a puzzling vertical gradient. The results from the Venera spectrophotometers (Moroz et al., 1980; Moroz 1983; Young et al., 1984) indicated mixing ratios that decreased from ~200 ppm at ~50 km to ~20 ppm near the surface. In a subsequent reanalysis of the Venera spectrophotometer results (Ignatiev et al., 1997), however, the data were found to be consistent with mixing ratios in the range 30±10 ppm, with evidence for a minimum at 10-20 km altitudes.

A new approach to the determination of the $H_2O$ content of the lower atmosphere was made possible by the discovery of near-infrared windows in the Venus nightside spectrum (Allen and Crawford 1984; Allen, 1987; Crisp et al., 1991). These windows are a series of wavelength regions between 0.9 and 2.5 μm in which the atmospheric absorption is sufficiently weak that thermal emission from the deep atmosphere or surface escapes to space. This enables observation of the lower atmosphere by remote sensing from orbiting spacecraft or earth-based telescopes. Table 1 lists the main Venus atmospheric windows giving their approximate range in wavelength and wavenumber. Several of the windows cover regions in which $H_2O$ absorption features are present, enabling a determination of the water vapor mixing ratio. Moreover since the windows probe different depths in the Venus atmosphere it is possible to obtain information on the vertical distribution of $H_2O$. The 2.3 μm window measures altitudes around 35 km, the 1.74 μm window around 24 km, and the 1.18 µm window is sensitive to altitudes from the surface to around 16km (Pollack et al., 1993; Tsang et al., 2008)

A number of infrared remote sensing studies of the water vapor content of the Venus deep atmosphere have been obtained using ground-based telescopes and spacecraft as listed in table 2. These results give $H_2O$ mixing ratio in the lower atmosphere ranging from 26 to 50 ppm. To determine the $H_2O$ mixing ratio it is necessary to use line lists giving the spectral line parameters, which are then used in radiative transfer models to derive synthetic spectra that can be compared with the observations. When the first of these studies were carried out in the early 1990s there were no reliable high temperature $H_2O$ line lists available. Thus many of the previous studies have used versions of the HITRAN or GEISA lists for the $H_2O$ line parameters. However, these lists are designed primarily for Earth atmosphere transmission studies and are not intended to be high temperature line databases. It is not clear how well they represent the $H_2O$ spectrum at the temperatures of up to ~700K encountered in the Venus deep atmosphere. There are also known to be significant systematic errors in the line intensities in some versions of the HITRAN database for certain spectral regions (e.g. Belmiloud et al., 2000; Casanova et al., 2006).

There have been considerable recent theoretical and experimental developments that improve our understanding of the $H_2O$ spectrum. In this paper an intercomparison is made of a number of $H_2O$ line lists with particular reference to the modeling of the Venus deep atmosphere. The results are used to provide recommendations on how best to model the Venus spectrum, and are used to make a reassessment of past measurements of the $H_2O$ mixing ratio.

## 2  Line lists and modeling methods

The line lists that were used in this analysis are listed in table 3. All these lists cover wavelengths ranges from the microwave or far-IR through to visible wavelengths, so the number of lines listed is a guide to their relative completeness.

## 2.1 HITRAN

HITRAN (**hi**gh resolution **tran**smission) is a large compilation of molecular spectral line data. The latest version, HITRAN 2004 (Rothman et al., 2005) includes data for around 1.7 million line transitions for 39 molecules. HITRAN evolved from the AFGL atmospheric absorption line compilation (McClatchey et al., 1973, Rothman et al., 1982). HITRAN has been through a series of updated versions. The main sources of near-infrared $H_2O$ line positions and intensities are summarized in table 4. HITRAN-86 (Rothman et al., 1987) used near-infrared $H_2O$ line parameters from the AFGL compilation of 1980 or earlier (Rothman 1981). In HITRAN 91/92 (Rothman et al., 1992) lines in the 8000-11500 $cm^{-1}$ region were updated using data from Mandin et al. (1988) and Chevillard et al. (1989). In HITRAN-96 (Rothman et al., 1998) lines in the 5750-7965 $cm^{-1}$ region were updated from the work of Toth (1994). HITRAN 2000 (Rothman et al., 2003) included a correction to the line intensities to lines at wavenumber above 8000 $cm^{-1}$. Giver et al. (2000) pointed out that these line intensities had been incorrectly converted from the values in the original sources in the previous HITRAN editions of 1992 and 1996. HITRAN 2000 also updated the line parameters in the 0.94 µm region based on the work of Brown et al. (2002). HITRAN 2004 (Rothman et al., 2005) includes a major update to all $H_2O$ line parameters from 500-8000 $cm^{-1}$ based on the work of Toth. This leaves the region from 8000-9500 $cm^{-1}$ as the only part of the near-infrared water vapor spectrum to be based primarily on older sources (in this case the work of Mandin et al., 1988). This region, however, covers the Venus 1.18 µm window that provides the deepest probe of the Venus atmosphere. Two editions of HITRAN were included in the analysis in this paper, HITRAN 2000 and HITRAN 2004.

## 2.2 GEISA

GEISA (Gestion et Etude des Informations Spectroscopiques Atmospheriques) is another large compilation of molecular spectral line data similar to HITRAN. The latest edition is GEISA 2003 (Jacquinet-Husson et al., 2008). GEISA, like HITRAN, is a compilation of data from other published sources and hence corresponding editions of HITRAN and GEISA often have very similar data. The sources of $H_2O$ line parameters used in GEISA are listed in table 4. For the $H_2O$ near-infrared line parameters the GEISA-97 edition (Jacquinet-Husson et al., 1999) is based on the same principal sources as HITRAN-96. GEISA-03 is similar to HITRAN 2000 except for the 8000-9500 cm$^{-1}$ region. The line intensities in this wavelength region in the GEISA-03 database have not been corrected for the errors which Giver et al. (2000) pointed out were present in both HITRAN and GEISA. Only HITRAN has incorporated the correction (in versions from HITRAN-2000 onwards). The GEISA-03 line intensities can be brought into agreement with HITRAN by applying the correction factor recommended by Giver et al. (2000). GEISA-03 does not include the major update to H$_2$O parameters from 500-8000 cm$^{-1}$ based on the work of Toth, that went into HITRAN 2004. The GEISA-97 and GEISA-03 editions are included in the analysis in this paper.

## 2.3  Preliminary High Temperature Database

This line list was provided by David Crisp and is included in this analysis as it was the H$_2$O line list used for the work of Meadows and Crisp (1996). It is a preliminary version of a high-temperature list derived by David Schwenke dating from around 1993, but differs from the PS list described below. It is referred to as High-T in the discussion here.

## 2.4  HITEMP

This is a high temperature H$_2$O list that was formerly released with the HITRAN database version of 1996 (Rothman et al., 1995). HITEMP is no longer provided on the HITRAN website, but is available as a commercial product from ONTAR corporation (www.ontar.com). The HITEMP H$_2$O list includes lines to an intensity limit of 3 x 10$^{-27}$ cm molecule$^{-1}$ at 1000 K. HITEMP is designed to be compatible with HITRAN-96 so lines common to both lists have identical data. The HITEMP release also includes CO$_2$ and CO line lists, though the CO$_2$ list has too high an intensity cutoff  (3 × 10$^{-27}$ cm molecule$^{-1}$ at 1000K) to be useful for modeling the Venus lower atmosphere. Note that in the Venus community, the name HITEMP has sometimes been used to refer to other high temperature line lists, in particular the CO$_2$ list due to Wattson described in Pollack et al. (1993), which has become the standard CO$_2$ line list for modelling Venus.

## 2.5 Partridge and Schwenke (PS)

The Partridge and Schwenke (1997) list is computed from an ab-initio potential energy surface (PES) adjusted to fit HITRAN line position data, combined with an ab-initio dipole moment surface (DMS). The list is based on 170,625 rovibrational energy levels and can be used up to very high temperatures (~4000 K). The version of the PS list used in this analysis was taken from the web site of Robert Kurucz (http://cfaku5.cfa.harvard.edu).

## 2.6 Barber et al. (BT2)

The BT2 line list (Barber et al., 2006) is a line list computed from the PES of Shirin et al. (2003) and the DMS of Schwenke and Partridge (2000). Both of these are improvements on the versions used by PS. It contains more than 500 million lines for 221,097 energy levels. Barber et al. (2006) show that the BT2 energy levels are in better agreement with experimental levels than those of PS. Schwenke and Partridge (2000) show that their DMS (used for the BT2 list) gives line intensities in better agreement with experiment than the DMS used by PS.

## 2.7 Radiative Transfer Model (VSTAR)

Synthetic spectra were calculated using the radiative transfer model VSTAR (Versatile Software for Transfer of Atmospheric Radiation; Bailey, 2006). VSTAR models the radiative transfer in a planetary atmosphere using line-by-line calculations of the molecular absorption, combined with scattering from aerosols, clouds and molecules. The multiple scattering solution of the radiative transfer equation is obtained using the discrete ordinate method as implemented in the DISORT software (Stamnes et al., 1988). Some examples of previous applications of VSTAR to the spectra of planetary atmospheres can be found in Hough et al., 2006; Bailey et al., 2007 and Bailey et al., 2008. While VSTAR can be used to model complete planetary atmospheres, for many of the tests described here, it was used to calculate the transmission of simple slab models, which are represented in VSTAR as atmospheres with a single homogenous layer, and no scatterers.

The optical depths for each layer of an atmosphere, at each wavelength, were calculated from the line parameters as described in Appendix A of Rothman et al. (1998). The total internal partition sums were calculated using the subroutine described by Fischer et al. (2003). The line shape was modeled as an approximated Voigt profile (Humlicek, 1982; Schreier, 1992) in the line core, and as a Van Vleck–Weisskopf profile (Van Vleck and Weisskopf, 1945) in the wings. The Van Vleck–Weisskopf profile is equivalent to a Lorentzian profile unless the line width is very large. The pressure broadened line width for the profile was, for $H_2O$, derived from the model described in section 2.8.

HITRAN, GEISA and similarly formatted lists such as HITEMP provide line intensities ($S_{HIT}$) in units of cm molecule$^{-1}$ at a reference temperature of 296 K. BT2 provides intensities in the form of Einstein A-coefficients ($A_{21}$). The BT2 intensities were converted to HITRAN form using (Simeckova et al., 2006)

$$S_{HIT} = \frac{g_s(2J+1)A_{21}}{8\pi c \nu_0^2 Q_{tot}(T_0)} e^{-hcE_1/kT_0}(1 - e^{-hc\nu_0/kT_0})$$

where J is the angular momentum quantum number for the upper state, and $g_s$ is its nuclear spin degeneracy (3 for ortho states, 1 for para states), $\nu_0$ is the line frequency (cm$^{-1}$), $E_1$ is the lower state energy (cm$^{-1}$) and $Q_{tot}(T_0)$ is the total internal partition sum at the reference temperature ($T_0$ = 296 K).

The Kurucz version of the PS list gives line intensities in the form log($gf$) where $g$ is the degeneracy factor and $f$ is the oscillator strength. These were converted to HITRAN intensities using

$$S_{HIT} = \frac{\pi g f q_e^2}{m_e c^2 Q_{tot}(T_0)} e^{-hcE_1/kT_0}(1 - e^{-hc\nu_0/kT_0})$$

where $q_e$ is the electron charge, $m_e$ is the electron mass and other symbols are as above.

## 2.8 Line Width Model

The HITRAN/GEISA type files also include parameters used to derive the pressure broadened line width needed for the Lorentzian (or Van Vleck–Weisskopf) line profile. These are provided for each line in the form of an air-broadened line width coefficient ($\gamma_{air}$ in cm$^{-1}$ atm$^{-1}$), a temperature exponent (*n*) for the air-broadened width, and a self-broadened width ($\gamma_{self}$). These data have evolved through different versions, with a complete overhaul of the $H_2O$ line width parameters being included in HITRAN 2004. The BT2 and PS lists do not include any line width data.

To ensure the use of a consistent line width model when comparing the different line lists, and to provide a line width appropriate for broadening in a $CO_2$ (rather than air) atmosphere, the HITRAN/GEISA line width parameters were ignored and line widths were derived from the following model for all line lists.

Brown et al. (2007) have shown that the $CO_2$ broadened width for $H_2O$ lines cannot be adequately represented by a simple scaling of the air-broadened width by a constant factor as has been used in some previous studies (Crisp et al., 1991; Pollack et al., 1993). The foreign broadened line widths used for this study were based on the data calculated by Delaye et al. (1989). These calculations have been found to agree well with measured broadening parameters (Langlois et al., 1994) in the 1.4 μm region over temperatures from 300-1200 K. Delaye et al. (1989) lists line widths and temperature coefficients for a range of values of the rotational quantum numbers (J, Ka, Kc). For a few transitions with rotational quantum numbers that were not listed in this tabulation, or for which the full set of rotational quantum numbers was not available (which is the case for some levels included in BT2), an averaged representation of the tabulated data as a function of J was used. A constant width (0.0271 $cm^{-1}$ $atm^{-1}$) and temperature exponent (0.30) was used for J>16. The self broadened widths, which have minimal effect at the small mixing ratios encountered in the Venus atmosphere, were based on averaged values from the data compiled by Robert Toth (http://mark4sun.jpl.nasa.gov/data/spec/H2O) and were assumed to be a function of J only.

## 2.9 Water Vapor Isotopologues

All the analysis in this paper concerns only the main isotopologue of water $H_2^{16}O$. Some, but not all, of the line lists in this study include lines of other isotopologues, but these were not included in the slab model comparisons described below. For Venus HDO is particularly important because the D/H ratio of Venus is more than 100 times higher than the terrestrial value making

HDO a significant absorber. Some of the same issues found for the main isotopologue may well also apply to HDO and will be investigated in a future study.

## 3 Line List Comparisons

### 3.1 Slab Models

To compare the representation of water vapor absorption in the various line lists, a set of models were computed of the transmission of homogenous slabs containing 30 ppm of water vapor. For line broadening purposes the $H_2O$ was assumed to be in an atmosphere containing pure $CO_2$, but no $CO_2$ absorption was included in the models. Models were computed for three wavelength regions (8000-9000 $cm^{-1}$; 5500-6000 $cm^{-1}$ and 4000–4500 $cm^{-1}$) covering the three main Venus nightside windows that have been used for $H_2O$ measurements (1.18 μm, 1.74 μm and 2.3 μm). For the first set of models the pressure and temperature were chosen to be approximately those in the deepest layers that contribute to Venus absorption at the selected wavelength, and the slab thickness was set at 10 km, a little less than one atmospheric scale height. Table 5 lists the parameters of these models.

Figure 1 shows a section of the 8000-9000 $cm^{-1}$ model that includes the 1.18 μm Venus window. In this region large differences are apparent between the various line lists. Over most of this region the high-T list has less absorption than any of the other lists. The HITEMP, BT2 and PS lists show the most absorption, and the HITRAN and GEISA lists are intermediate. Figures 2 and 3 show a similar comparison for sections of the 5500-6000 $cm^{-1}$ model (1.74 μm window) and the 4000-4500 $cm^{-1}$ model (2.3 μm window). Again differences between the line lists can be seen, although they are generally smaller than those apparent at 1.18 μm. In most cases the HITEMP, BT2 and PS lists show more absorption than the HITRAN and GEISA lists. The differences are much greater at some wavelengths than others (e.g. at around 1.741 μm).

From the models calculated at actual Venus pressures, the large amount of pressure broadening makes it difficult to see what is causing the apparent differences. A second set of models was therefore calculated with the pressure reduced to 1 bar ($10^5$ Pa), but the slab thickness increased correspondingly to keep the total column density of $H_2O$ roughly the same. The parameters of these models are also given in table 5. At this pressure individual lines are much more clearly resolved. Sections of the transmission spectra are shown in figures 4, 5 and 6 for three of the line lists, HITRAN 2004, HITEMP and BT2 (generally GEISA is similar to HITRAN, and PS is similar to HITEMP and BT2). It is apparent from these comparisons that there are often lines in HITEMP and BT2 that are not present in HITRAN. Two strong examples are marked with dotted lines in figure 4. These two example lines have intensities of $1.4 \times 10^{-23}$ and $2.3 \times 10^{-23}$ cm molecule$^{-1}$ at 700K. These extra lines are generally "hot" lines that become relatively strong at Venus lower atmosphere temperatures (500-700 K), but would not be significant at the normal Earth atmosphere temperatures for which HITRAN and GEISA are designed. There are also occasions of lines in HITRAN that are not in BT2 although this is uncommon. An example is shown in figure 4 at around 1.1699 μm.

Figures 4, 5 and 6 also show that there are often differences in the line positions in BT2 compared with HITRAN (and HITEMP, which is based on HITRAN for strong lines). This reflects the fact that there are significant uncertainties in computed line positions, as compared with the measured values included in HITRAN. Barber et al., 2006, state that in BT2 48.7% of energy levels are within 0.1 cm$^{-1}$ of measured values and 91.4% are within 0.3 cm$^{-1}$. In the regions modelled here line positions in BT2 often disagree with those in HITRAN by up to about 0.2 cm$^{-1}$.

## 3.2 Line Intensity Histograms

To further investigate the completeness of the various line lists for hot lines at Venus lower atmosphere temperatures, histograms of line intensities were plotted over the range of each of the slab models described in the previous section. One histogram used the line intensities at the actual temperature used for the slab models (700 K for 8000-9000 cm$^{-1}$ [1.18 μm window], 570 K for 5500-6000 cm$^{-1}$ [1.74 μm window] and 500 K for 4000-4500 cm$^{-1}$ [2.3 μm window]). These are the approximate temperatures of the Venus atmosphere layers sampled by the corresponding windows. The second histogram was plotted for line intensities at a temperature of 300 K for each wavelength region. The results are shown in figures 7, 8 and 9. Figures 8 and 9 show that all line lists produce very similar histograms at 300 K, at least down to intensities of about $10^{-25}$ cm molecule$^{-1}$. However, the high temperature histograms show an obvious difference between the standard HITRAN and GEISA lists, and the other high temperature lists. HITRAN and GEISA are clearly substantially incomplete for lines with intensities below about $10^{-23}$ cm molecule$^{-1}$ at temperatures of 500 K and 570 K. These are lines that are strong enough to show substantial absorption in Venus atmosphere conditions as seen in figures 4, 5 and 6. For example, the two "missing" lines marked in figure 4 have intensities of about $10^{-23}$ cm molecule$^{-1}$. In the 2.3 µm region (figure 9) there is also a significant difference between HITRAN and GEISA, with GEISA falling significantly below HITRAN in the histograms indicating it is less complete.

In the 8000-9000 cm$^{-1}$ region (1.18 µm window) the incompleteness of HITRAN and GEISA at 700 K is also readily apparent. However, in this case the histogram also shows HITRAN and GEISA falling below the other lists even at 300 K for intensities below $10^{-24}$ cm molecule$^{-1}$. This indicates the completeness of the HITRAN and GEISA data in this wavelength region is significantly poorer than at the longer wavelengths.

It is therefore apparent that HITRAN and GEISA (all versions) are not sufficiently complete in high temperature lines to be used to model the Venus lower atmosphere at temperatures of 500-700 K. This should not be surprising. These databases were not intended for high temperature use. Nevertheless they have been widely used for past Venus atmosphere models.

### 3.3 Intensities of Strong Lines

While the incompleteness of HITRAN and GEISA for hot lines accounts for some of the differences between line lists apparent in figures 1-3, it is not the only effect. Comparison of model spectra at low temperatures shows that there are also significant differences in line intensities even for strong lines that are in all line lists and important at all temperatures. These effects can be investigated by a direct comparison of the tabulated line intensities at the standard 296 K temperature (or appropriately converted intensities as described in section 2.7 for BT2 and PS). For this comparison BT2 was used as a reference, as its line intensities are computed in a consistent way (rather than compiled from many different sources as is the case for HITRAN and GEISA), and are believed to be more accurate than earlier lists such as PS (Barber et al., 2006; Schwenke and Partridge, 2000).

Figure 10 shows the comparison of HITRAN 2004, HITRAN 2000 and GEISA-97 (which is equivalent to HITRAN-96) with BT2. Lines have been identified as the same line in the two lists if they have the same quantum numbers and a line position and lower state energy within 1 cm$^{-1}$. The ratio of the line intensity with that in BT2 is plotted over a wide range of wavelength for all lines with intensities greater than $3 \times 10^{-24}$ cm molecule$^{-1}$.

Examination of figure 10 shows that as updates have been included in HITRAN and GEISA the agreement of line intensities with BT2 has improved. In GEISA-97 (HITRAN-96) most of the near-infrared line intensities fall below the BT2 values by 10 to 25%. In HITRAN 2000 the 10200-11200 cm$^{-1}$ region was updated based on a very thorough study by Brown et al. (2002), and the scatter in line intensities is much reduced, and the data come into excellent agreement with BT2. Changes in the 8200-9200 cm$^{-1}$ region are also apparent as a result of the correction to line intensities according to Giver et al. (2000). In HITRAN 2004 the region from 500-8000 cm$^{-1}$ was updated. The improvement over the 5000-8000 cm$^{-1}$ range is clearly shown in figure 10 with these regions coming into better agreement with BT2. There is also some improvement in agreement for the 3000-4500 cm$^{-1}$ region, although here there is still a lot of scatter, and the HITRAN intensities are mostly below those in BT2. The 8200-9200 cm$^{-1}$ region (1.18 μm window) was not updated in HITRAN 2004, and shows the poorest agreement with BT2 with intensities mostly 5-20% below those in BT2. The fact that the inclusion of improved experimental data always seems to bring HITRAN into better agreement with BT2, strongly suggests that the line intensities in this list are excellent and supports its authors claim that the Schwenke and Partridge DMS on which its intensities are based "is the most accurate in existence" (Barber et al., 2006).

Figure 10 also suggests that while HITRAN 2004 is a substantial improvement on previous versions, its line intensities seem to be poorest in two wavelength regions that are particularly important for Venus, those containing the 2.3 μm and 1.18 μm windows. It also shows that all earlier versions of HITRAN and GEISA (including GEISA 2003 which is similar to HITRAN 2000) appear to significantly underestimate line intensities over most of the near-IR range. Systematic errors in HITRAN $H_2O$ near-IR line intensities, in particular in the 8000-9500 $cm^{-1}$ range, have been reported in a number of other studies. Belmiloud et al. (2000) report that measured integrated intensities in the $2\nu + \delta$ polyad (around 8800 $cm^{-1}$) are 26% higher than those derived from HITRAN-96. Smith et al. (2004) report that the same band is underestimated by 20% in HITRAN 2000, and also find a 15% underestimation for the 1.4 μm band. Casanova et al. (2006) use measurements of solar radiation reaching the Earth's surface to determine that line intensities in the 8000-9500 $cm^{-1}$ region are underestimated by 18% in HITRAN 2004 as compared with lines in the 3000-8000 $cm^{-1}$ range.

Figure 11 shows a similar comparison with BT2 for strong lines in the high temperature line lists (high-T, PS and HITEMP). The high-T list shows large discrepancies with BT2, the line intensities generally being substantially lower. One of the largest differences is seen in the 8500 $cm^{-1}$ region where intensities fall to 50-60% of those in BT2. This is the region of the 1.18 μm window where this list was used for the Meadows and Crisp (1996) study of Venus. The PS list shows quite good agreement with BT2 for lines <4000 $cm^{-1}$, but substantial differences at higher frequencies. As discussed by Schwenke and Partridge (2000) and Barber et al. (2006), the PS list is based on an older dipole moment surface, and the improved version used for BT2 is believed to give much better line intensities.

The HITEMP list shows the same pattern in figure 11 as that for GEISA-97 in figure 10. This reflects the fact that HITEMP uses the HITRAN-96 data for lines common to both lists (i.e. all the strong low temperature lines), and HITRAN-96 is essentially the same as GEISA-97 for these lines. This means that HITEMP includes all the systematic errors previously discussed for early HITRAN and GEISA versions. HITEMP includes the erroneous intensities of lines above 8000 $cm^{-1}$ noted by Giver et al. (2000).

### 3.4 Summary of Line Lists

The suitability of the various $H_2O$ line lists for modeling the Venus deep atmosphere can be summarized as follows.

HITRAN and GEISA (all versions) are deficient in hot lines important at temperatures encountered in the Venus lower atmosphere. They also show significant systematic errors in the intensities of strong lines, particularly, for the 8000–9500 cm$^{-1}$ region (corresponding to the 1.18 μm window). These line lists should not be used for modeling Venus.

The PS and high-T lists contain hot lines, but the line intensities are not as accurate as those in the more recent BT2 list. BT2 should be used in preference.

HITEMP contains hot lines appropriate for the temperatures encountered in the Venus atmosphere. However, for strong lines, HITEMP is identical to HITRAN 96, and thus contains the systematic errors present in this HITRAN version, as well as lacking the improvements made in more recent HITRAN/GEISA updates.

BT2 is complete for hot lines, and contains accurate line intensities. BT2 is recommended as currently the best $H_2O$ line list for modeling the Venus deep atmosphere. The main deficiency of BT2 is that line positions are generally not as accurate as those measured experimentally. At the relatively low spectral resolving power of most Venus observations this is not likely to be a serious problem.

## 4  Corrections to Past Measurements

Meadows and Crisp (1996, hereafter MC) based their analysis of Venus on the high-T line list, which is shown in figure 11 to have large line intensity errors in the 1.18 μm region. To investigate the effect of these errors on the derived water vapor abundance, models of the Venus spectrum for the 1.18 μm window were calculated using the high-T and BT2 line lists. The spectra were calculated using VSTAR, following a modelling procedure that aimed, as far as possible, to reproduce that described by MC. The temperature structure was based on the VIRA temperature profile (Seiff et al., 1985) . Clouds were modeled using four modes of $H_2SO_4$ particles as described by Crisp (1986). $CO_2$ line parameters were taken from the high-temperature database of Wattson (Pollack et al., 1993) and a sub-Lorentzian line shape model was used with the line wings calculated up to large distances from the core, with no additional $CO_2$ continuum being included as described by MC. Line parameters for molecules other than $CO_2$ and $H_2O$ were taken from HITRAN 2004. The radiative transfer solution for each spectral point was obtained using an eight stream discrete ordinate method using the DISORT solver (Stamnes et al., 1988).

Models were calculated for the high-T line list at the $H_2O$ mixing ratio of 45ppm derived by MC, and for the BT2 list at a series of $H_2O$ mixing ratios from 0–45 ppm in one ppm steps. MC derived their water vapor mixing ratio from the ratio of the intensity for a wavelength range of 1.1698–1.1769 μm (on the water vapor absorbed edge of the 1.18 μm window) to that for the wavelength range 1.1840–1.1868 mm (near the peak of the window where water vapor absorption is not significant). From the models calculated here, the BT2 line list would lead to the same value for this intensity ratio for a $H_2O$ mixing ratio of 27 ppm as is given by the high-T list for a mixing ratio of 45 ppm. It is therefore concluded that the MC water vapor mixing ratio of 45±10 ppm, should actually be 27±6 ppm when derived using the more complete and accurate BT2 line list. This correction brings the MC value into better agreement with other estimates listed in table 2.

The correction to the MC water vapor abundance is straightforward to derive because it results from large systematic errors in the line intensities, and the wavelengths used to derive the water vapor mixing ratio were clearly specified. For other studies that used versions of the HITRAN and GEISA databases, corrections may also be necessary, but they cannot be straightforwardly estimated, because they will result from a combination of two effects, the incompleteness of the line lists, and the systematic errors for strong lines. The magnitude of any correction will depend on the precise spectral region used for the study and the details of the fitting procedure. The comparisons in figures 1-3 suggest that the corrections are likely to be smaller than those needed for the MC study.

With recent studies yielding $H_2O$ mixing ratios of high precision (±2 ppm for Marcq et al., 2008) it is clear that the accuracy and completeness of the water vapor line list is crucial, and it is important that such observations are reanalysed using the best available modern high temperature line lists. Using line lists that accurately represent hot lines may also enable more detailed studies of the water vapor profile in the Venus atmosphere when spectra of sufficient resolution are analysed. The hot lines will originate primarily in the lower and hotter part of the atmosphere, whereas other lines will include contributions from all altitudes. The 1.18 μm window includes hot lines that could be clearly separated from other lines at spectral resolving powers of a few thousand.

## 5   Conclusions

A comparison of eight water vapor spectral line lists reveals significant discrepancies between near infrared $H_2O$ absorption derived from different lists for conditions appropriate to the Venus lower atmosphere (temperatures up to ~700 K). The differences are shown to be a combination of two factors. Line lists such as HITRAN and GEISA that are not designed for high temperature use, are deficient in "hot" lines that become important at the 500–700 K temperatures encountered in the Venus atmosphere. Several line lists show evidence of significant systematic errors in line intensities, even for strong low temperature lines.

The Meadows and Crisp (1996) study of the Venus lower atmosphere derived a $H_2O$ mixing ratio of 45±10 ppm, based on a preliminary high temperature line list that had large systematic errors in line intensity. When these errors are corrected the $H_2O$ mixing ratio drops to 27±6 ppm. Other studies that used versions of the HITRAN and GEISA line lists may also somewhat overestimate the $H_2O$ mixing ratio. The BT2 line list (Barber et al., 2006) is recommended as the most complete and accurate current representation of the $H_2O$ spectrum at Venus temperatures.

## Acknowledgments

The author thanks Bruno Bezard, David Crisp, Nicole Jacquinet, Larry Rothman and Victoria Meadows for discussions and provision of spectral line data.

**Figure Captions**

Figure 1 – Transmission of a 10 km thick slab containing 30 ppm of water vapor at a temperature of 700 K and pressure of 90 bar ($9 \times 10^6$ Pa). The conditions correspond to those near the surface of Venus that are probed by night-side observations in the 1.18 μm window. Large differences between the six line lists can be seen. In this wavelength region GEISA 97 is the same as GEISA 2003, and HITRAN 2000 is the same as HITRAN 2004.

Figure 2 — Transmission of a 10 km thick slab containing 30 ppm of water vapor at a temperature of 570 K and a pressure of 20 bar ($2 \times 10^6$ Pa). The conditions correspond to those at an altitude of ~20 km in the Venus atmosphere that are probed by night-side observations in the 1.74 μm window. Significant differences between the five line lists can be seen. In this wavelength region GEISA 97 and HITRAN 2000 are the same as GEISA 2003.

Figure 3 — Transmission of a 10 km thick slab containing 30 ppm of water vapor at a temperature of 500 K and a pressure of 10 bar ($10^6$ Pa). The conditions correspond to those at an altitude of ~30 km in the Venus atmosphere that are probed by night-side observations in the 2.3 μm window. Significant differences between the five line lists can be seen. In this wavelength region GEISA 97 and HITRAN 2000 are the same as GEISA 2003.

Figure 4 — Transmission of a 1000 km thick slab containing 30 ppm of water vapor at a temperature of 700 K and a pressure of 1 bar ($10^5$ Pa) calculated with HITRAN 2004, and with two high temperature line lists, HITEMP and BT2. The wavelength range is on the edge of the 1.18 μm Venus window. The high temperature line lists include a number of lines not included in HITRAN 2004. Two of the stronger examples are marked with dotted lines. The dashed line indicates an unusual case of a line that is in HITRAN and HITEMP but not in BT2.

Figure 5 — Transmission of a 200 km thick slab containing 30 ppm of water vapor at a temperature of 570 K and a pressure of 1 bar ($10^5$ Pa) calculated with HITRAN 2004, and with two high temperature line lists, HITEMP and BT2. The wavelength range is in the 1.74 μm

Venus window. The high temperature line lists include a number of lines not included in HITRAN 2004.

Figure 6 — Transmission of a 100 km thick slab containing 30 ppm of water vapor at a temperature of 500 K and a pressure of 1 bar ($10^5$ Pa) calculated with HITRAN 2004, and with two high temperature line lists, HITEMP and BT2. The wavelength range is in the 2.3 μm Venus window. The high temperature line lists include a number of lines not included in HITRAN 2004.

Figure 7 — Histogram showing number of $H_2O$ lines as a function of line intensity for the 8000–9000 $cm^{-1}$ region (containing the 1.18 μm Venus window) at 700 K and 300 K. At 300 K the HITRAN and GEISA lists are complete to intensities of about $10^{-24}$ whereas at 700 K HITRAN and GEISA are missing lines for intensities of $10^{-23}$ and lower as compared with the high temperature line lists.

Figure 8 — Histogram showing number of $H_2O$ lines as a function of line intensity for the 5500–6000 $cm^{-1}$ region (containing the 1.74 μm Venus window) at 570 K and 300 K. At 300 K the HITRAN and GEISA lists are complete to intensities of about $10^{-25}$ whereas at 570 K HITRAN and GEISA are missing lines for intensities of $10^{-23}$ and lower as compared with the high temperature line lists.

Figure 9 — Histogram showing number of $H_2O$ lines as a function of line intensity for the 4000–4500 $cm^{-1}$ region (containing the 2.3 μm Venus window) at 500 K and 300 K. At 300 K the HITRAN list is complete to intensities of about $10^{-25}$ whereas at 500 K HITRAN and GEISA are missing lines for intensities of $10^{-23}$ and lower as compared with the high temperature line lists. GEISA appears significantly lower than HITRAN in these histograms suggesting it is not as complete in this wavelength region.

Figure 10 — Ratio of line intensities at 296 K with those for the corresponding line in BT2 for GEISA 97 (equivalent to HITRAN-96), HITRAN 2000 and HITRAN 2004. Lines with intensities (in BT2) greater than $3 \times 10^{-24}$ cm $molecule^{-1}$ are plotted.

Figure 11 — Ratio of line intensities at 296 K with those for the corresponding line in BT2 for the high-T, PS and HITEMP lists. Lines with intensities (in BT2) greater than $3 \times 10^{-24}$ cm molecule$^{-1}$ are plotted.

Table 1 – The main Venus nightside atmospheric windows. Those in bold are the ones most used for H$_2$O abundance determination.

| Window Name | Wavelength range (μm) | Wavenumber range (cm$^{-1}$) |
|---|---|---|
| 1.00 μm | 0.96 — 1.04 | 9610 — 10410 |
| 1.10 μm | 1.07 — 1.11 | 9010 — 9350 |
| **1.18 μm** | **1.14 — 1.20** | **8330 — 8770** |
| 1.27 μm | 1.255 — 1.282 | 7800 — 7970 |
| 1.31 μm | 1.301 — 1.311 | 7630 — 7685 |
| **1.74 μm** | **1.690 — 1.755** | **5700 — 5920** |
| **2.3 μm** | **2.20 — 2.48** | **4030 — 4545** |

Table 2 – Remote sensing measurements of Venus H$_2$O mixing ratio

| Reference | Instrument | Window[c] | H$_2$O mixing ratio (ppm) | H$_2$O line-list used |
|---|---|---|---|---|
| Crisp et al., 1991 | AAT/FIGS | 2.3 / 1.27 µm | 40±20 | HITRAN-86 |
| Carlson et al., 1991 | Galileo/NIMS | 2.3 µm<br>1.74 µm | $25^{+25}_{-13}$<br>$50^{+50}_{-25}$ | HITRAN-86 |
| Pollack et al., 1993 | AAT/FIGS | 2.3 µm<br>1.74 µm | 30±6<br>30±7.5 | HITRAN-91[b] |
| Drossart et al., 1993 | Galileo/NIMS | 1.18 µm | 30±15 | GEISA-91[c] |
| De Bergh et al., 1995 | CFHT/FTS | 2.3 µm<br>1.74 µm<br>1.18 µm | $30^{+15}_{-10}$<br>30±10<br>30±15 | GEISA-91[c] |
| Meadows and Crisp, 1996 | AAT/IRIS | 1.18 µm | 45±10 | High-T |
| Marcq et al., 2006 | IRTF/SPEX | 2.3 µm | 26±4 | GEISA-97 |
| Marcq et al., 2008 | VEX/VIRTIS-H | 2.3 µm | 31±2 | GEISA-97 |
| Bezard et al., 2008 | VEX/VIRTIS-M | 1.18 µm | 32±7 | GEISA-97 |

Notes

a – The 2.3 µm window samples altitudes around 35 km, 1.74 µm about 24 km, and 1.18 µm from the surface to around 16km.

b – Pollack et al., 1993 also investigated a preliminary high temperature water vapor line list. However their results for H$_2$O were based on lines from HITRAN-91.

c – Bruno Bezard, private communication.

Table 3 – $H_2O$ line lists compared in this study

| Name | Number of $H_2O$ lines | Reference |
|---|---|---|
| GEISA 97 | 50,217 | Jacquinet-Husson et al., 1999 |
| HITRAN 2000 | 53,271 | Rothman et al., 2003 |
| GEISA 2003 | 58,726 | Jacquinet-Husson et al., 2008 |
| HITRAN 2004 | 63,196 | Rothman et al., 2005 |
| High-T | 270,888 | Meadows and Crisp, 1996 |
| HITEMP | 1,283,468 | Rothman et al., 1995 |
| Partridge & Schwenke | 65,912,356 | Partridge and Schwenke, 1997 |
| BT2 | 505,806,202 | Barber et al., 2006 |

Table 4 – Main sources of near-infrared H$_2$O line parameters in HITRAN and GEISA

| Wavenumber range | 2900-4300 | 5200-6200 | 6500-7600 | 8200-9300 | 10200-11300 |
|---|---|---|---|---|---|
| Venus windows | 2.3 µm | 1.74 µm | | 1.18 µm | |
| HITRAN-86 | Rothman 1981 or earlier | Rothman 1981 or earlier | Rothman 1981 or earlier | Rothman 1981 or earlier | Rothman 1981 or earlier |
| HITRAN 91/92 GEISA 92 | unchanged | unchanged | unchanged | Mandin et al., 1988 | Chevillard et al., 1989 |
| HITRAN 96 GEISA 97 | unchanged | Some lines updated (Toth, 1994) | Toth 1994 | unchanged | unchanged |
| HITRAN 2000 GEISA 2003 | unchanged | unchanged | unchanged | Intensity correction (Giver et al. 2000) in HITRAN but not in GEISA | Brown et al., 2002 |
| HITRAN 2004 | Toth[a] | Toth[a] | Toth[a] | unchanged | unchanged |

a - Toth, R.A., Linelists of water vapor parameters from 500 to 8000 cm$^{-1}$,
   http://mark4sun.jpl.nasa.gov/data/spec/H2O

Table 5 — Parameters of homogenous slab models.

| Wavenumber range (cm$^{-1}$) | Pressure (bar) | Temp (K) | Thickness (km) | Figure |
|---|---|---|---|---|
| 8000-9000 | 90 | 700 | 10 | Figure 1 |
| 5500-6000 | 20 | 570 | 10 | Figure 2 |
| 4000-4500 | 10 | 500 | 10 | Figure 3 |
| 8000-9000 | 1 | 700 | 1000 | Figure 4 |
| 5500-6000 | 1 | 570 | 200 | Figure 5 |
| 4000-4500 | 1 | 500 | 100 | Figure 6 |

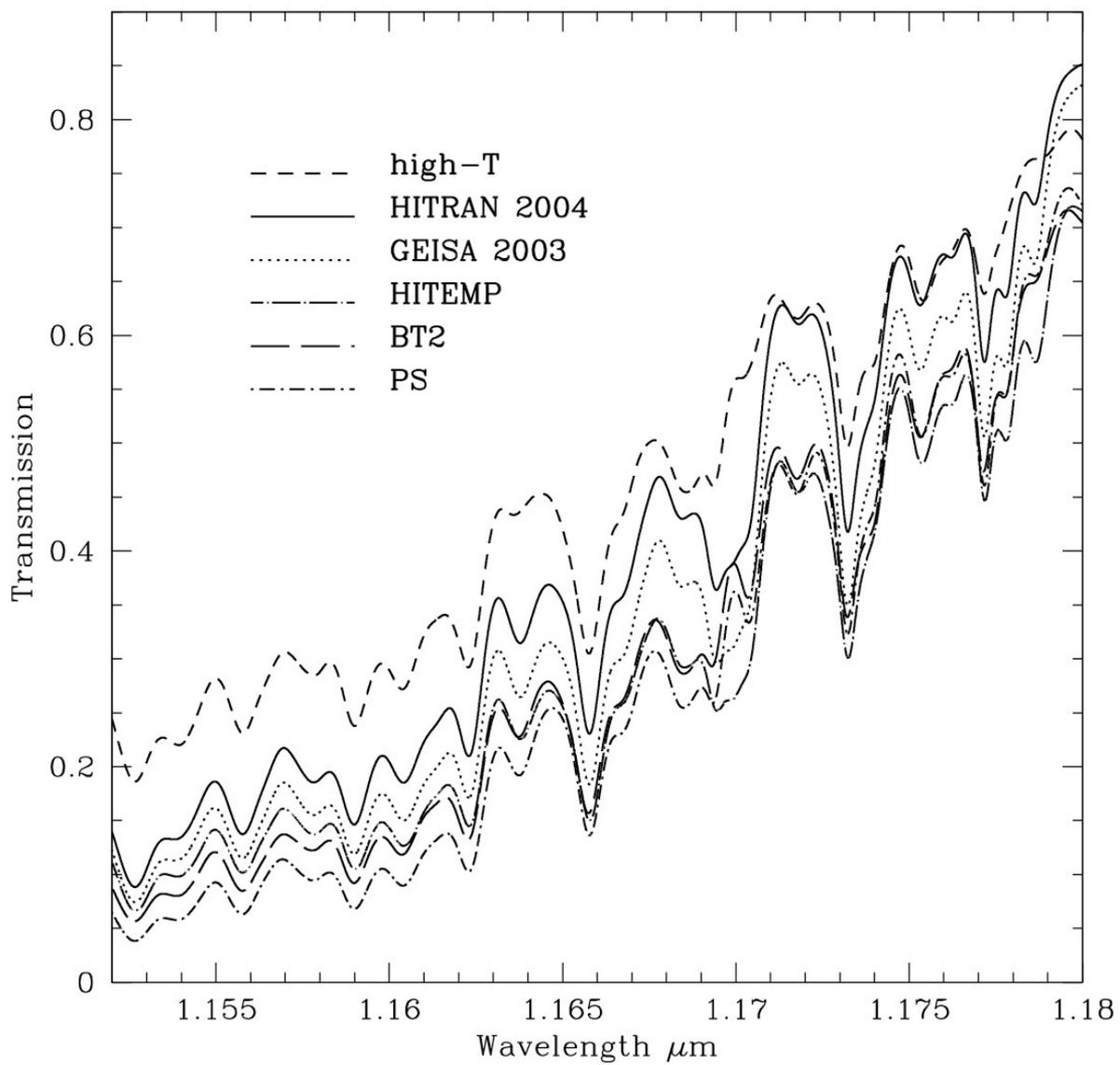

Figure 1 — Bailey, Water vapor line parameters

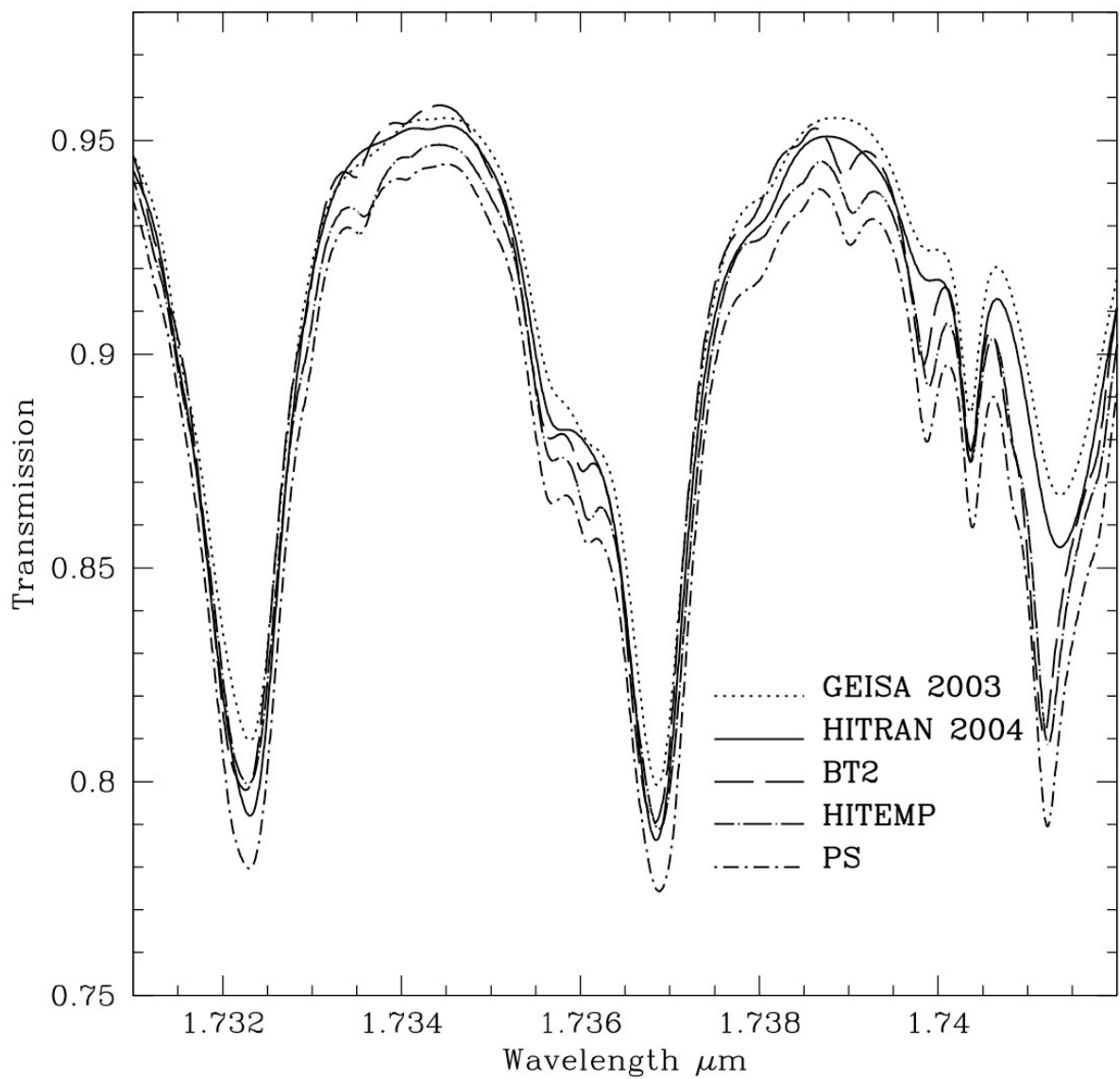

Figure 2 — Bailey, Water vapor line parameters

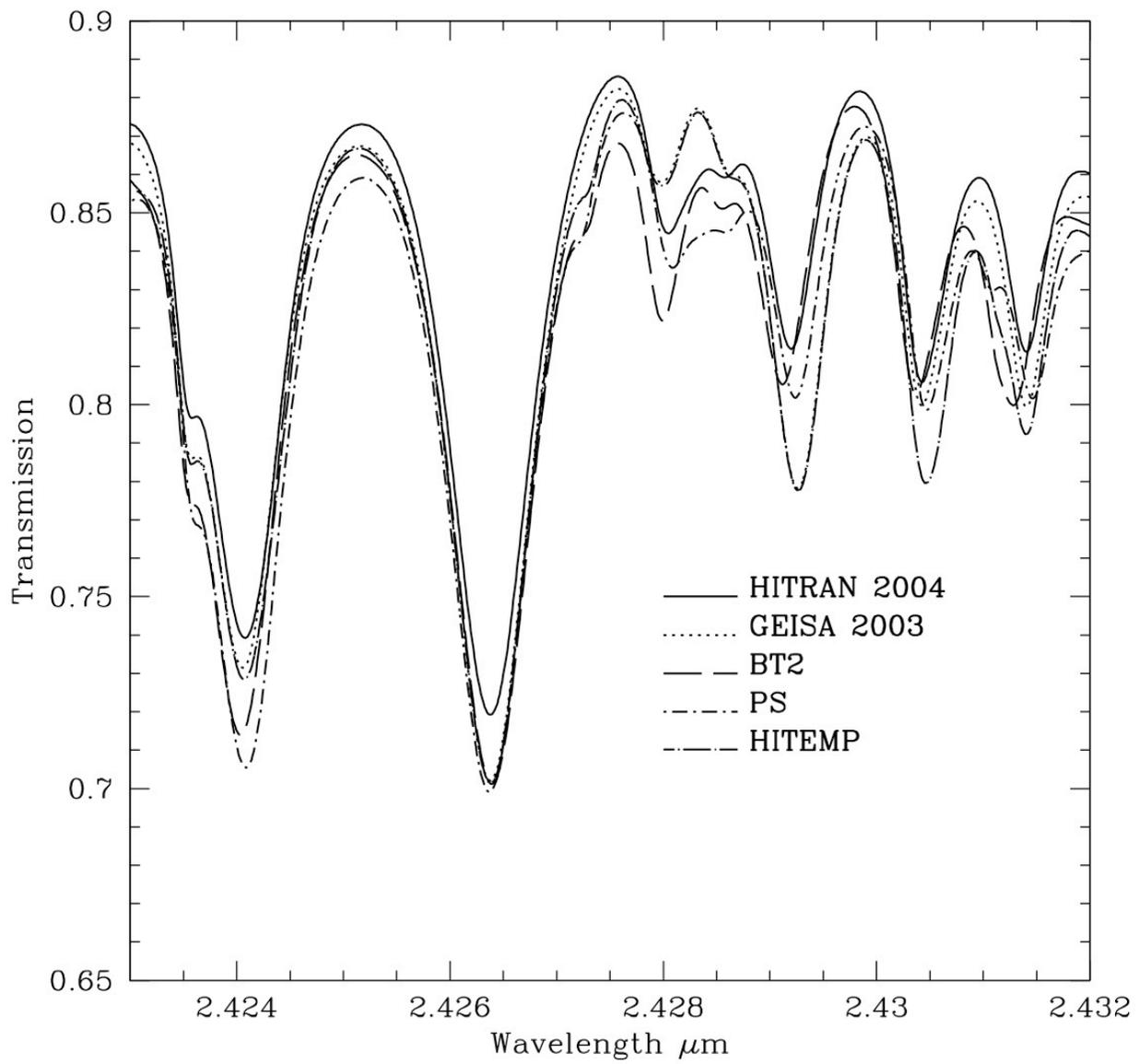

Figure 3 — Bailey, Water vapor line parameters

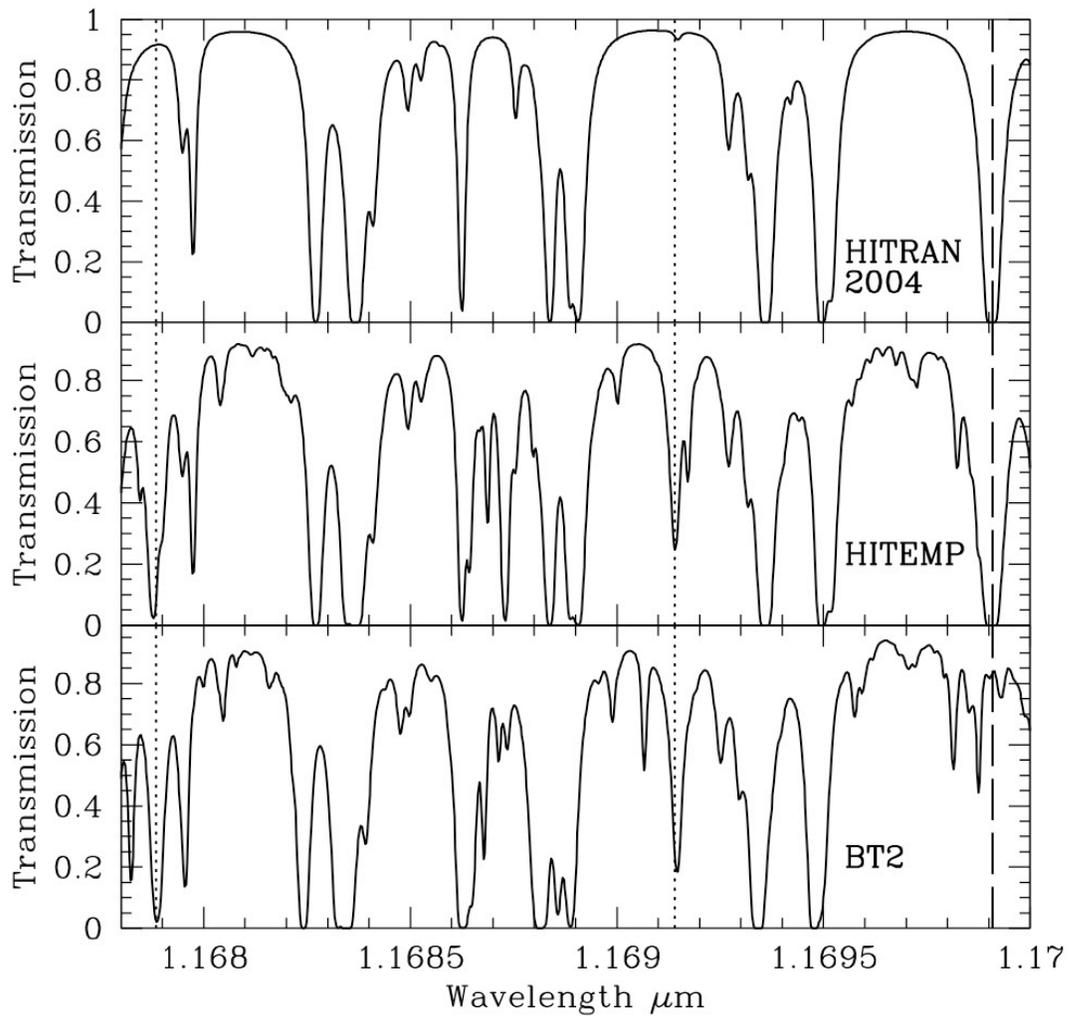

Figure 4 — Bailey, Water vapor line parameters

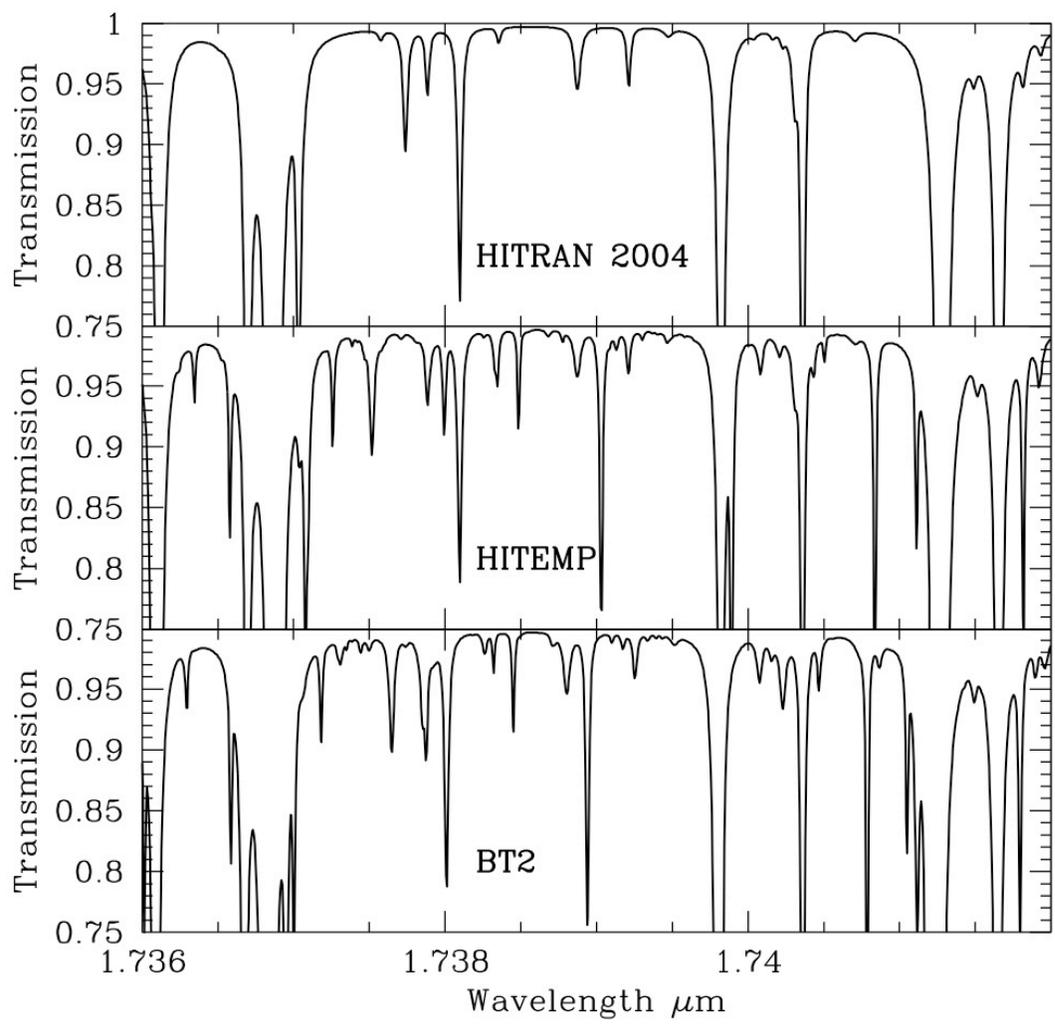

Figure 5 — Bailey, Water vapor line parameters

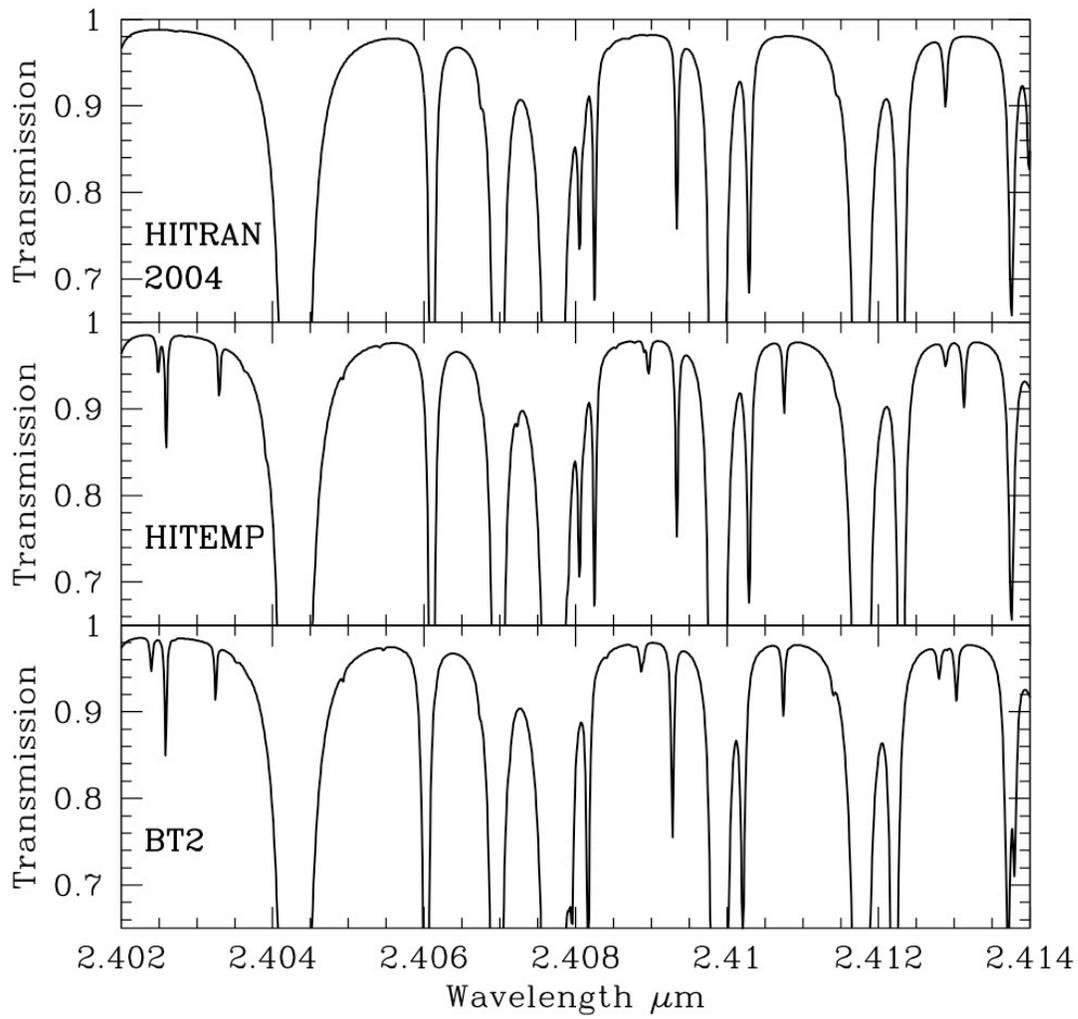

Figure 6 — Bailey, Water vapor line parameters

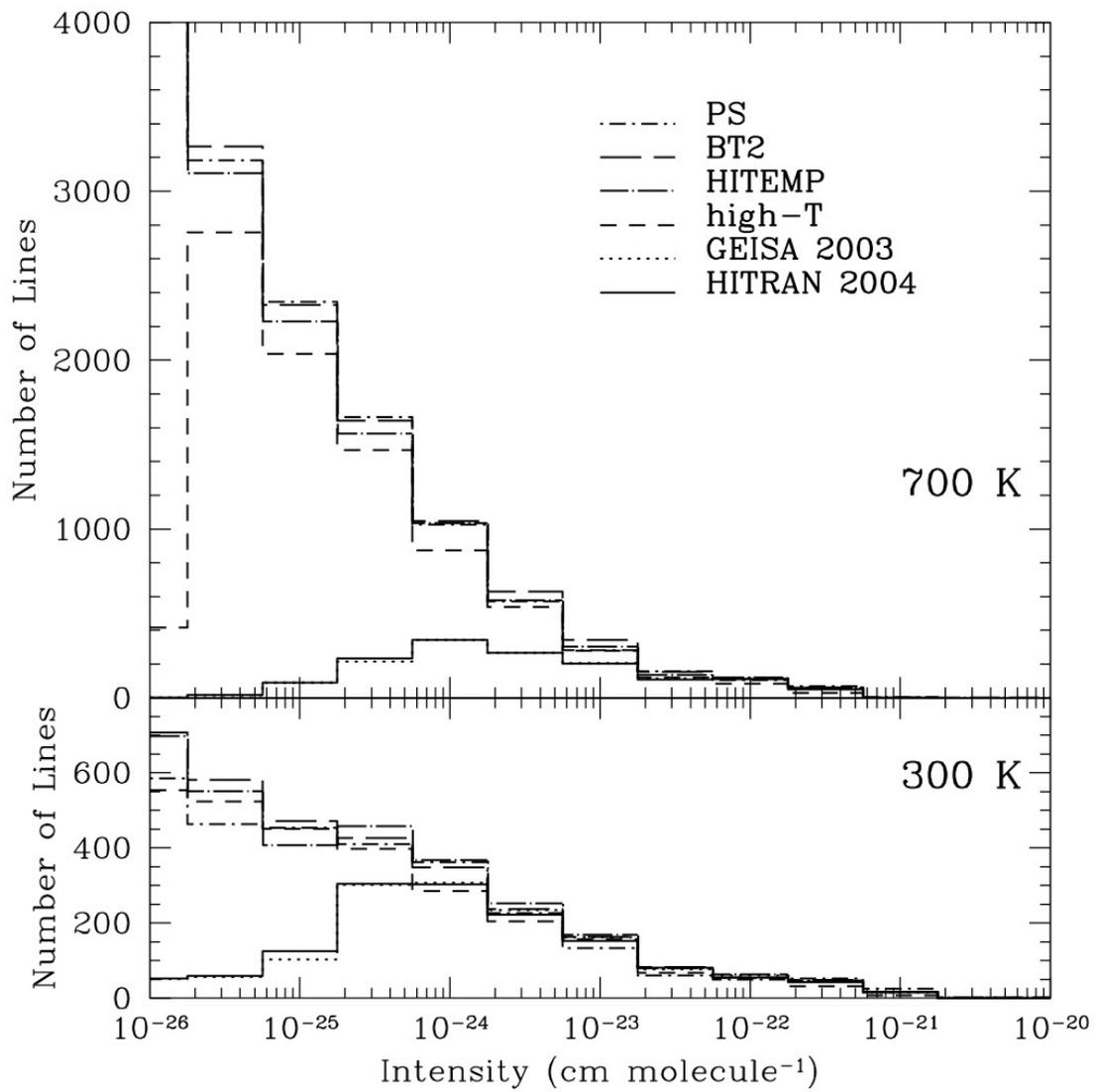

Figure 7 — Bailey, Water vapor line parameters

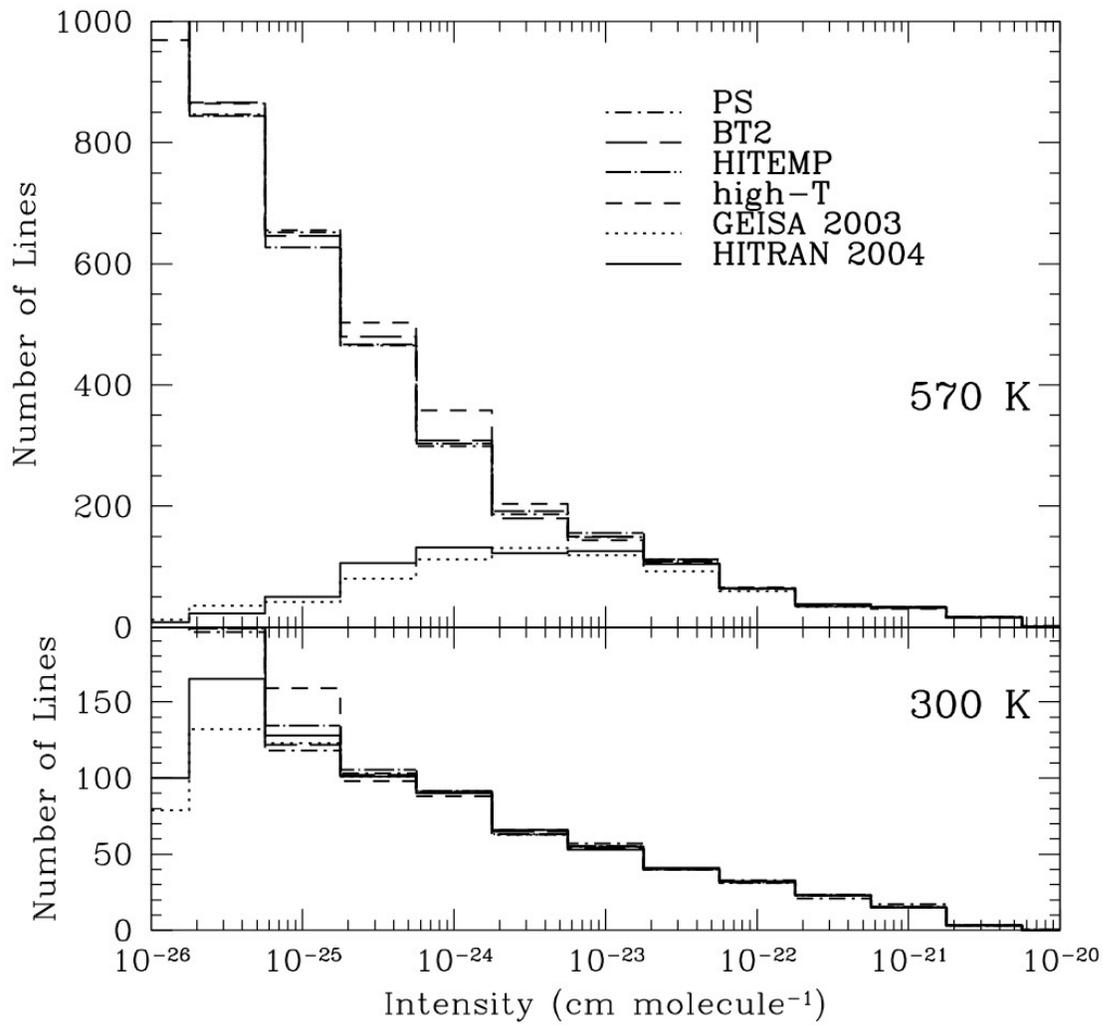

Figure 8 — Bailey, Water vapor line parameters

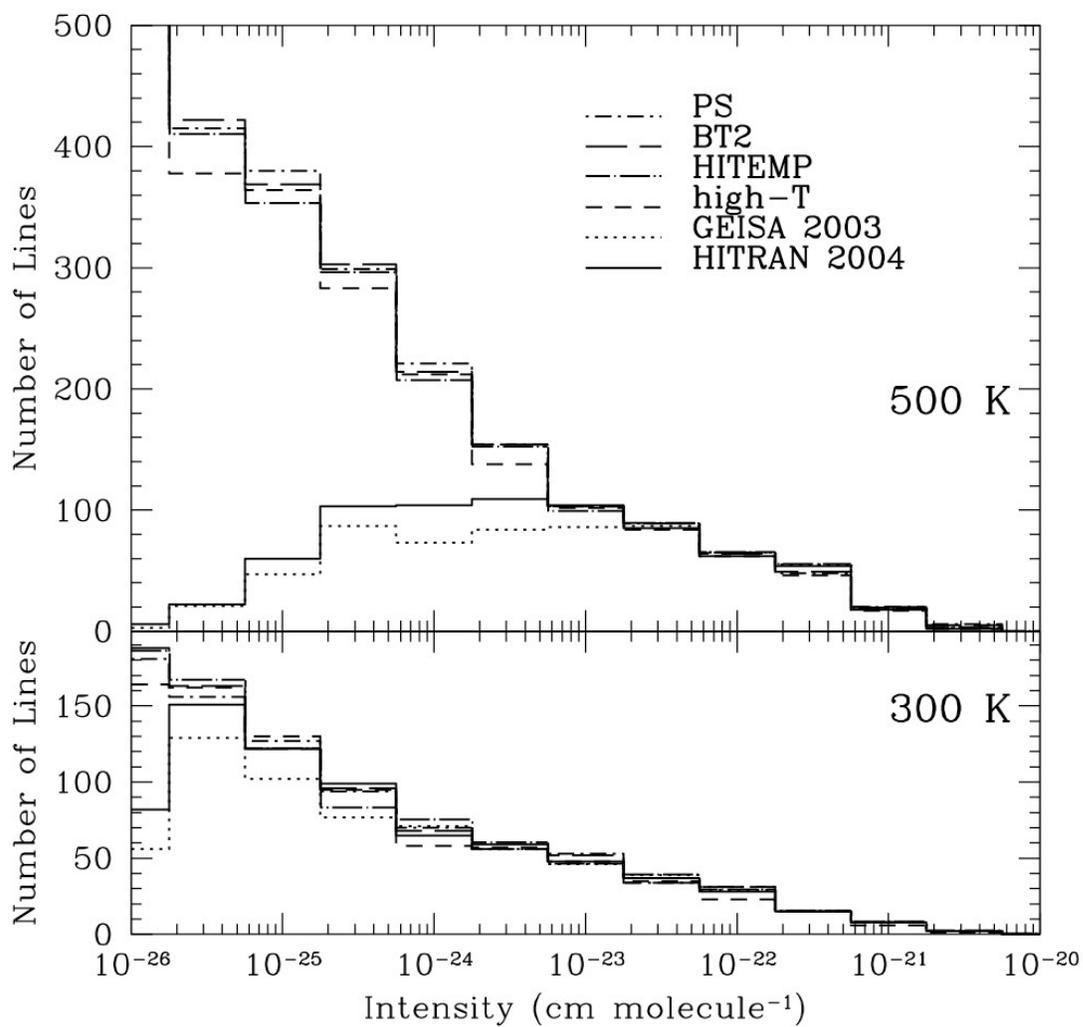

Figure 9 — Bailey, Water vapor line parameters

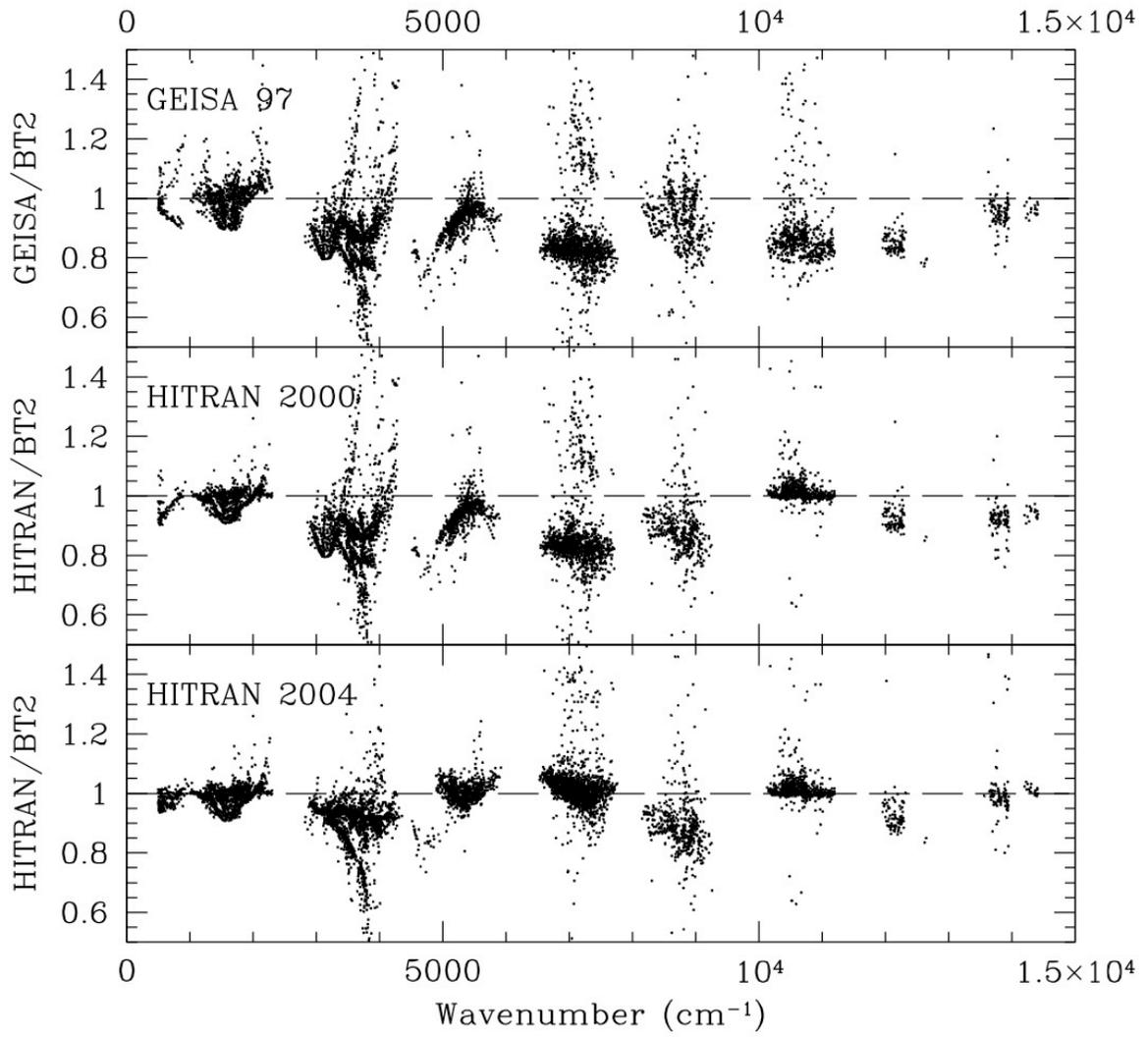

Figure 10 — Bailey, Water vapor line parameters

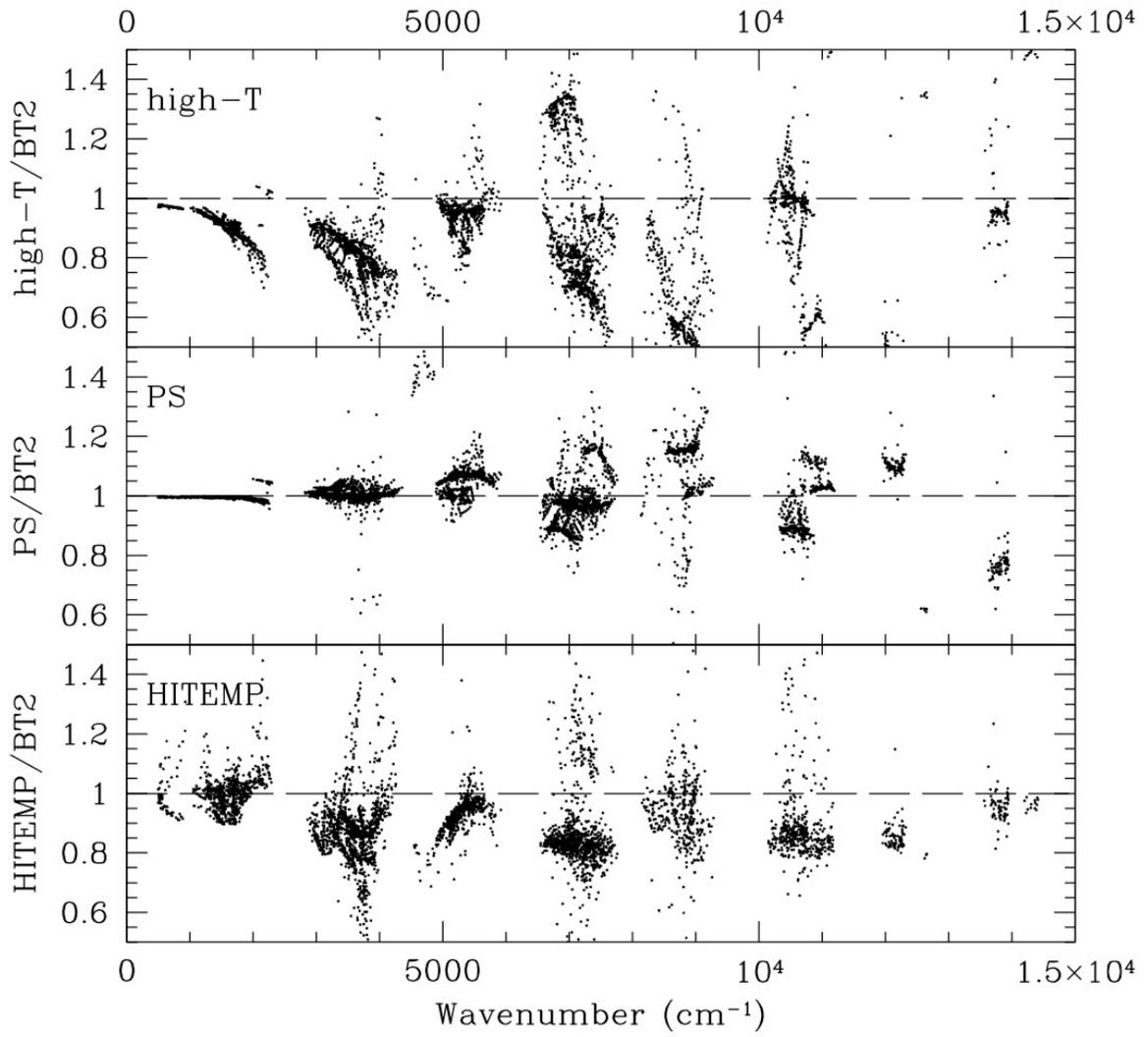

Figure 11 — Bailey, Water vapor line parameters